\documentclass{cernrep} 
\usepackage{texnames}
\usepackage[T1]{fontenc}
\usepackage[bookmarks, colorlinks=true, linktoc=page, linkcolor=red, citecolor=red, urlcolor=red]{hyperref}
\begin{document}
\title{LHC Run-2 and Future Prospects}
 
\author {J. T. Boyd}

\institute{CERN, Geneva, Switzerland}

\begin{abstract}
The lecture discusses both the current status of the LHC collider as well as its future running scenarios. 
In addition a selection of the latest physics results from ATLAS/CMS and LHCb is presented.
\end{abstract}

\keywords{LHC Run-2; EPSHEP2019.}

\maketitle 
 
\section{The LHC Collider}

The LHC is the highest energy particle collider in the world, situated in the 27 km long, former LEP tunnel at CERN. It is an extremely 
sophisticated machine, using super conducting 8T dipole magnets to steer the high-energy proton beams. The magnets are cooled to an 
operating temperature of 1.9K using super-fluid liquid helium. Given the extreme energy of the beams, the LHC needs a complex machine 
protection system, relying on a large number of beam instrumentation devices to monitor the beam position and beam losses.

The two key parameters for a collider are the collision energy, and the luminosity $L$ which is a measure of the number of collisions. The 
number of events for a specific process ($N$) is given by $N = \sigma \times L$, where $\sigma$ is the production cross-section for that 
process and $L$ is the integrated luminosity. 

The luminosity at a collider is given by the formula: $L = n_{\rm{b}} N_{\rm{1}} N_{\rm{2}} F/ 4 \pi \epsilon \beta^\ast$ and can be increased
 by increasing the number of protons per bunch ($N_{\rm{1}}, N_{\rm{2}}$), the number of colliding bunches ($n_{\rm{b}}$), or reducing the
 transverse size of the beam at the collision point  which can be done by using a lower emittance ($\epsilon$) beam, or by squeezing the 
beam more with the focussing magnets (reducing $\beta^\ast$). The crossing-angle between the beams, needed to avoid parasitic collisions due to the 
short distance between bunches, reduces the luminosity, and is encapsulated in the geometric factor $F$ in the equation.

The main machine parameters for the LHC are shown in Table~\ref{tab:LHCparams}, for the design, Run-1, Run-2, as well as the expectation for Run-3 and the 
High Luminosity upgrade (HL-LHC). It can be seen that all of the design parameters have been exceeded, except the collision energy, and the number of 
colliding bunches. The LHC experts have continually improved the running scenario to increase the luminosity, and during Run-2 the 
design luminosity of $10^{34}$ cm$^{-2}$s$^{-1}$, was achieved and surpassed by a factor of two at the end of Run-2. As well as improving the 
instantaneous luminosity, the availability of the machine was dramatically improved during Run-2 which led to a large dataset for physics;
 in 2016,2017,2018 running, the machine was providing physics collisions during 50\% of the allocated physics time, which is very 
impressive for a super conducting collider. An important parameter for the LHC experiments is the pileup, which is determined by the 
luminosity per bunch, and is a measure of the number of inelastic $pp$ interactions that occur per bunch crossing. Higher pileup gives 
more luminosity (for a fixed number of bunches) but makes physics analysis more difficult due to the signals in the detector from the 
additional interactions.  

The HL-LHC will operate from 2026 with the goal of delivering 3000 fb$^{-1}$ in the following ten years (an increase of a factor of ten 
compared to the expected dataset at that time). In order to achieve this the injector needs to be upgraded to provide higher intensity 
beam, the focussing magnets will be replaced to be able to squeeze the beam more, and various components will be upgraded to be able to 
cope with the increased radiation and stored energy. The pileup in ATLAS and CMS will increase significantly (to 150 interactions per 
bunch crossing) and the detectors will need large upgrades to be able to make physics measurements at this large pileup, as well as to 
be able to cope with the associated radiation.

\begin{table}[tbhp]
   \centering
  \begin{tabular}{|l|c|c|c|c|c|c|}
  \hline
Parameter & Design & Run-1 & Run-2 & Run-3 & HL-LHC \\
  \hline
Energy [TeV] & 14 & 7/8 & 13 & 14 & 14 \\
Bunch spacing [ns] & 25 & 50 & 25 & 25 & 25 \\
Bunch Intensity [$10^{11}$ ppb] & 1.15 & 1.6 & 1.2 & up to 1.8 & 2.2 \\
Number of bunches & 2800 & 1400 & 2500 & 2800 & 2800  \\
Emittance [$\mu$m] & 3.5 & 2.2 & 2.2 & 2.5 & 2.5 \\
$\beta^{*}$ [cm] & 55 & 80 & 30 $\to$ 25  & 30 $\to$ 25 & down to 15  \\
Crossing angle [$\mu$rad] & 285 & - & 300 $\to$ 260 & 300 $\to$ 260 & TBD  \\
Peak Luminosity [$10^{34}$ cm$^{-2}$s$^{-1}$] & 1.0 & 0.8 & 2.0 & 2.0 & 5.0  \\
Peak pileup & 25 & 45 & 60 & 55 & 150  \\
 \hline
  \end{tabular}
  \caption{Summary of main accelerator parameters for the LHC, showing the design values, and those used during Run-1 and Run-2, as well as the expected parameters for Run-3 and the HL-LHC.}
  \label{tab:LHCparams}
\end{table}

\section{Run-2 physics highlights and future prospects}
\subsection{The LHC detectors}
ATLAS and CMS are the general-purpose detectors at the LHC with the same physics goals. There are significant differences 
in the detector designs, but despite these they have very similar physics performance.

The main differences in the detectors relate to the magnet design, ATLAS utilizes a 2~T solenoid to provide the magnetic field to bend charged 
particles in the central detector region, with three toroidal magnets (one barrel toroid and two endcaps) to bend muons in the muon spectrometer. CMS 
uses a single large solenoid with field of 3.8~T for both of these roles. Following on from this the ATLAS calorimeters are placed outside the 
thin solenoid, whereas in CMS the calorimeters are inside the solenoid.

\subsection{The Run-2 dataset}
\begin{figure}[htb]
   \centering
   \includegraphics[width=52mm]{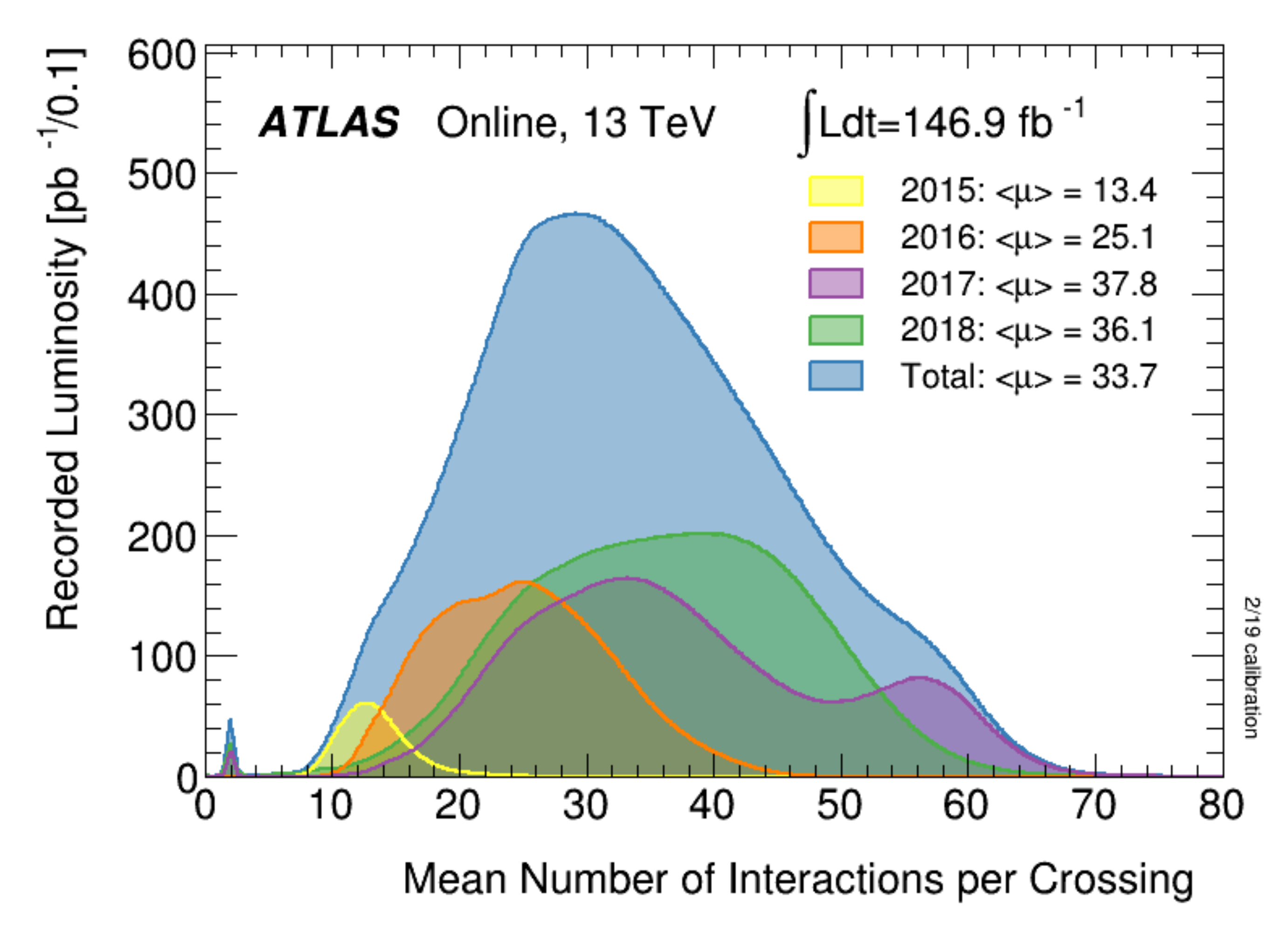}
   \includegraphics[width=52mm]{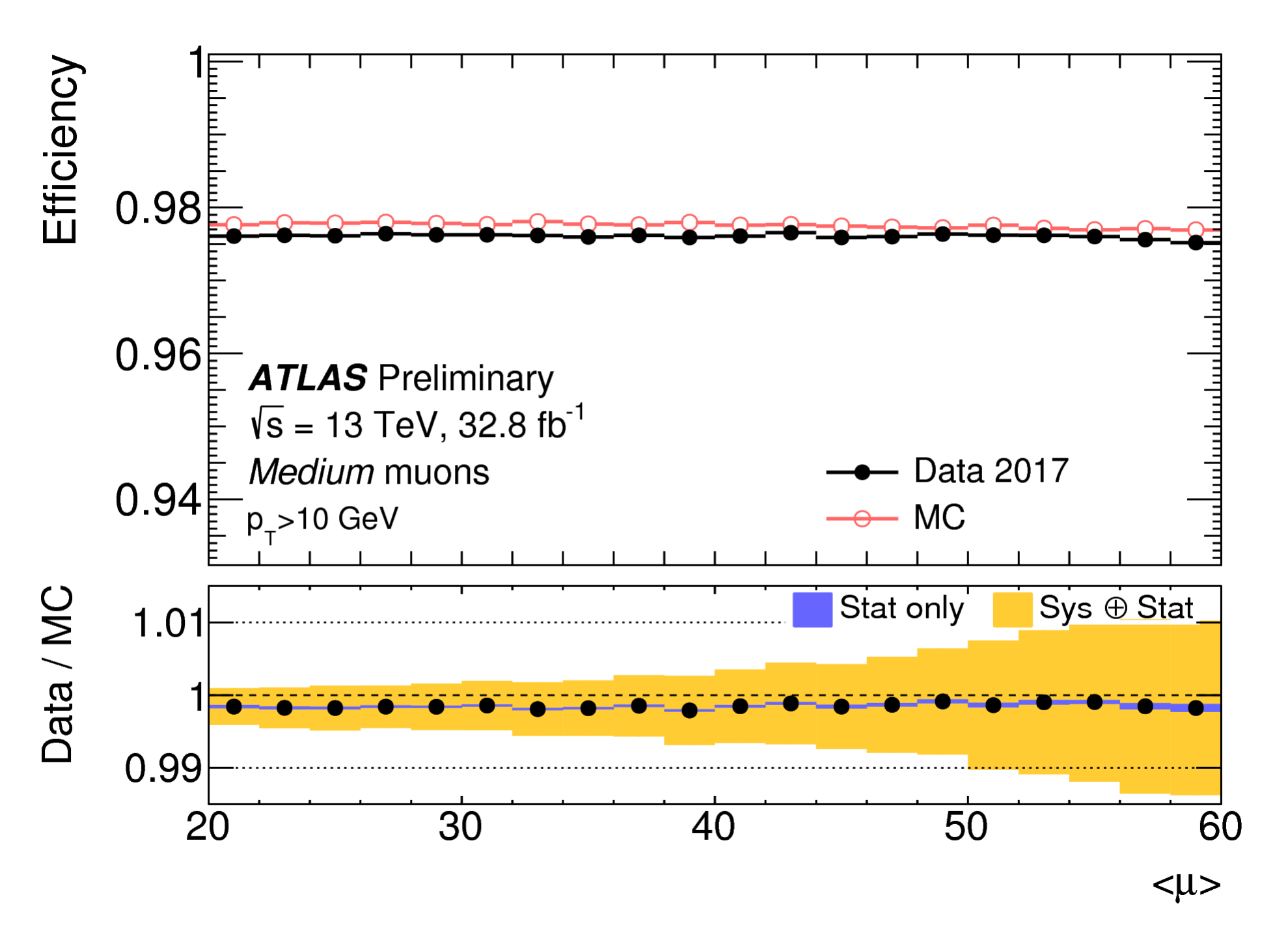}
   \includegraphics[width=52mm]{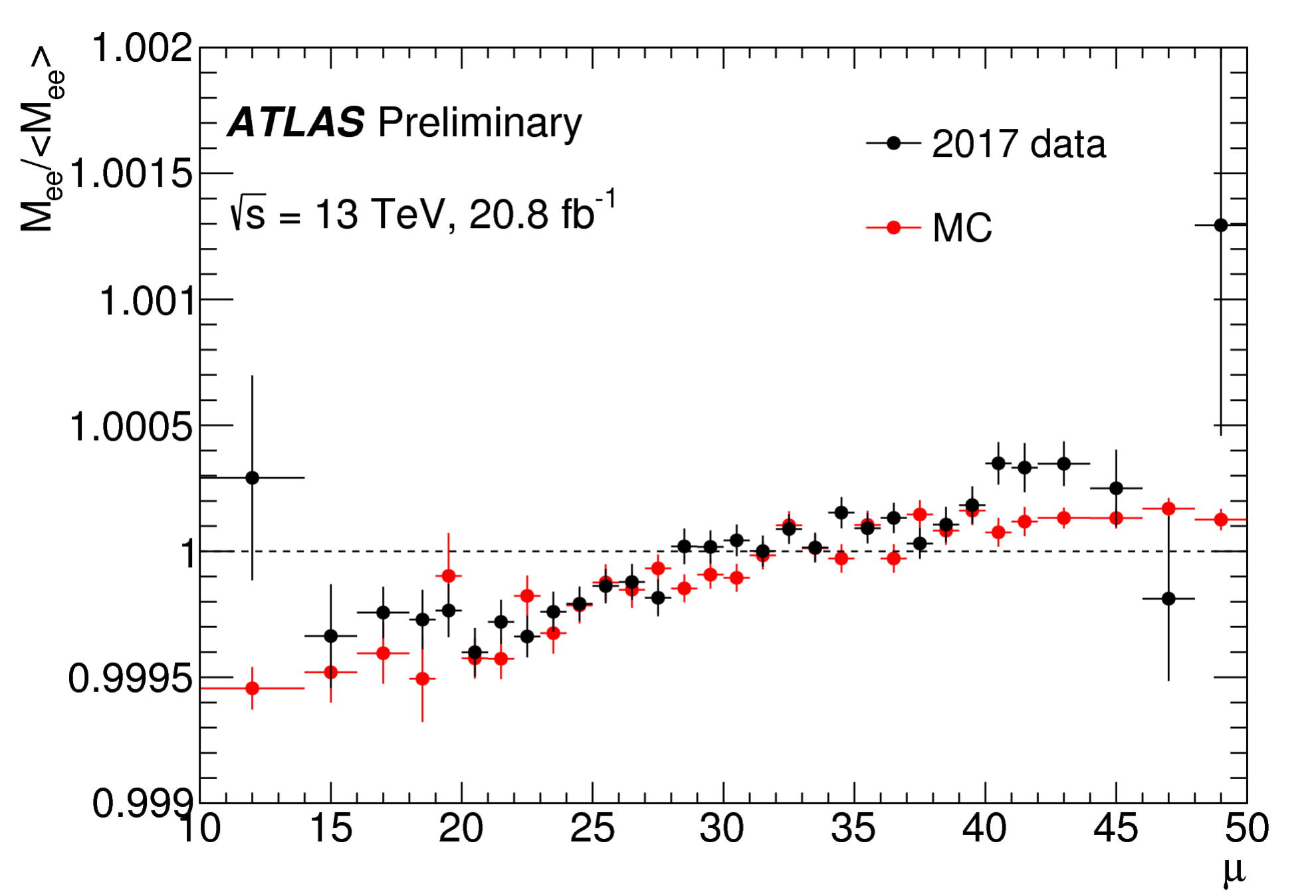}
   \caption{(left) The Run-2 pileup distribution. An example of the pileup robustness of (middle) the reconstructed muon efficiency and (right) the electron energy scale.}
   \label{fig:pileup}
\end{figure}

During Run-2 ATLAS and CMS collected $\approx$140 fb$^{-1}$ of data for physics analysis. The cost of this large dataset was the significantly larger 
pileup compared to the design value of 25, as can be seen in Figure~\ref{fig:pileup}, and it has been one of the main challenges for the experiments to be able 
to efficiently trigger, reconstruct the data, and carry out physics analyses with the Run-2 pileup. The experiments have put a huge 
effort into updating the trigger and reconstruction algorithms to improve the performance at high pileup. Figure~\ref{fig:pileup} shows examples of the 
achieved pileup robustness in terms of reconstructed particles both for efficiency and resolution.

\subsection{Higgs physics}
\begin{figure}[htb]
   \centering
   \includegraphics[width=65mm]{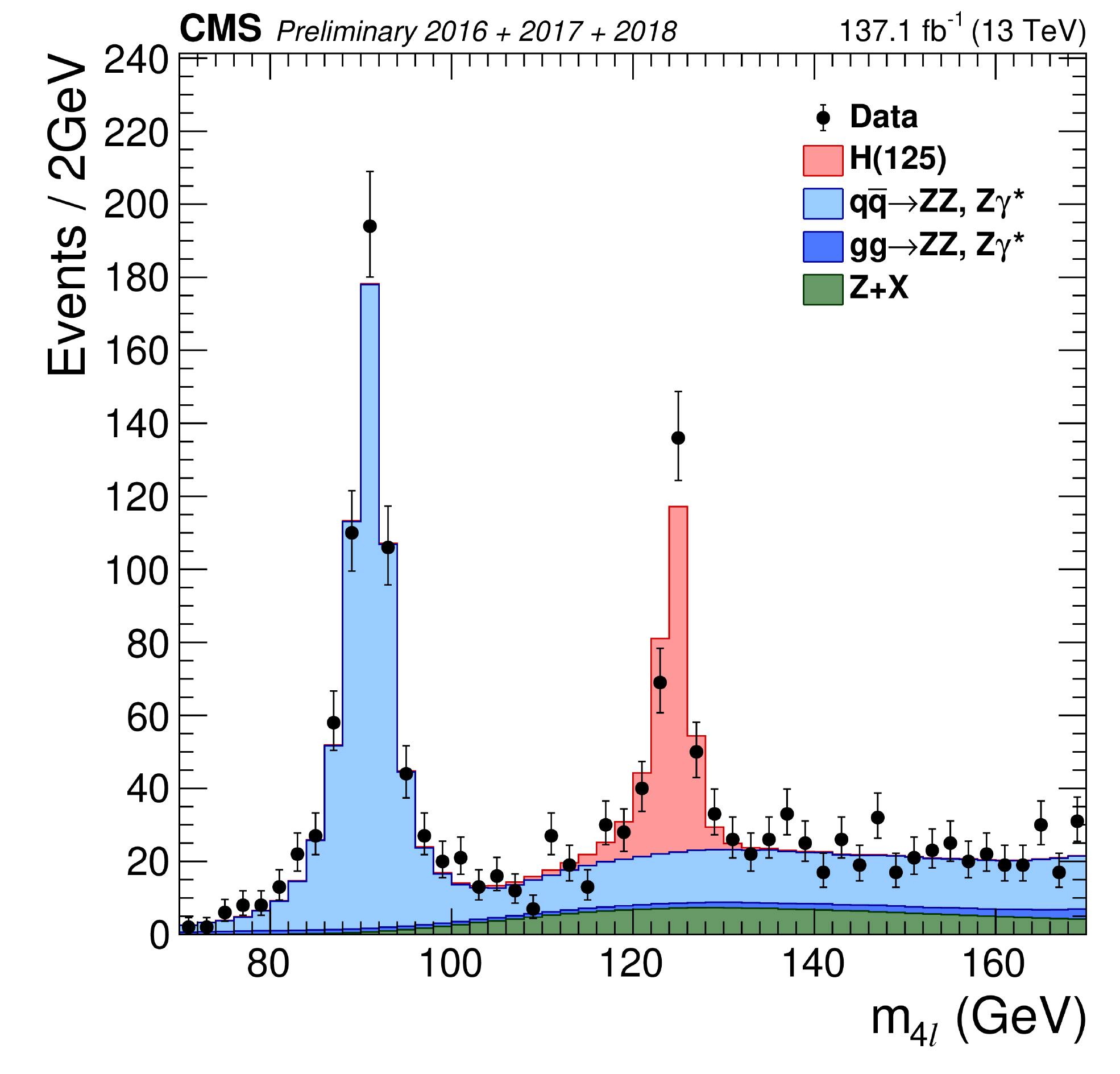}
   \includegraphics[width=85mm]{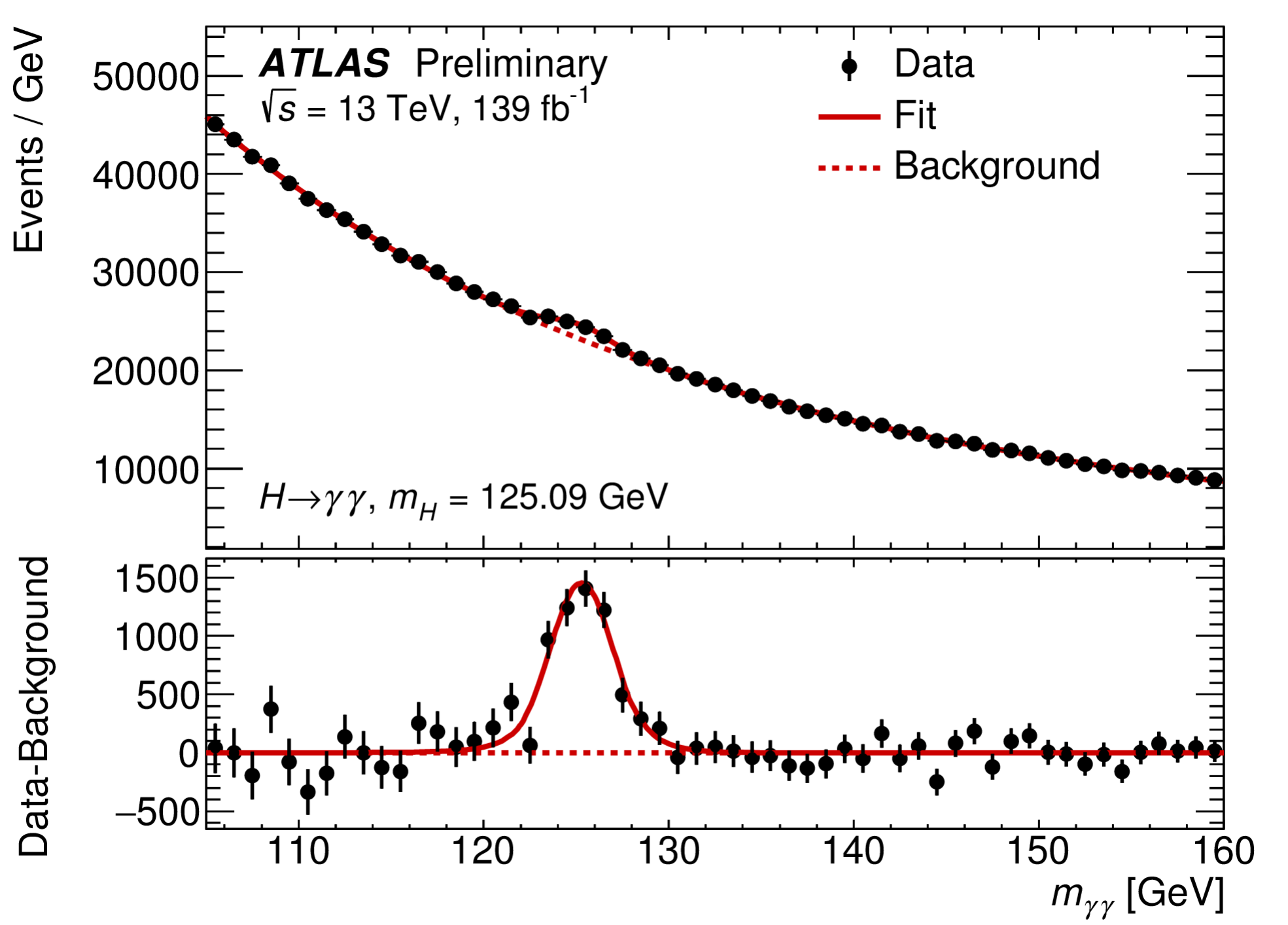}
   \caption{Reconstructed Higgs candidate mass distributions in the $H \to \gamma \gamma$ (left) and  $H \to 4\ell$ (right) channels.}
   \label{fig:higgs1}
\end{figure}

The main modes for Higgs 
production at the LHC are (in order of decreasing cross-section): gluon-fusion (ggF), Vector-Boson fusion (VBF), production in 
association with a vector boson (VH) and production in association with a pair of top-quarks (ttH). In VBF production the scattered 
quarks are likely to form forward jets on the two sides of the detector, which can be used to tag such events. The main decay modes for the Higgs 
boson are shown in Table~\ref{tab:HiggsDecays}. Experimentally the modes with the best mass resolution are important, as this allows to separate the signal 
from the background in a much more reliable way. The $H \to \gamma \gamma$ and $H \to Z Z^\ast \to 4\ell$ ($\ell = e/\mu$) both have excellent 
mass resolution of $\approx$1-2\%, and these were the modes used for the Higgs discovery in 2012, despite the fact they have very low branching fractions (BF). 
Figure~\ref{fig:higgs1} shows the mass distributions
 for these two channels for the full Run-2 dataset. For $H \to \gamma \gamma$ the signal to background (S/B) is low, but the total number of selected Higgs 
events is a few thousand, whereas for $H \to 4\ell$ the S/B is high, but the total number of signal events is an order of magnitude less.

\begin{table}[tbhp]
   \centering
  \begin{tabular}{|l|c||l|c|}
  \hline
  \multicolumn{2}{|c||}{ Poor mass resolution channels} & \multicolumn{2}{c|}{ Good mass resolution channels} \\
Decay Mode & BF (\%)  & Decay Mode & BF (\%)  \\
  \hline
$H \to b\overline{b}$ & 58.2 & $H \to ZZ^\ast$ & 2.6 (0.012 $e,\mu$)  \\
$H \to WW^\ast$ & 21.4 (1.1 $e,\mu$)   & $H \to \gamma \gamma$ & 0.23  \\
$H \to gg$ & 8.2 &  $H \to Z \gamma$ & 0.15 (0.008 $e,\mu$)  \\
$H \to \tau^+\tau^-$ & 6.3  &  $H \to \mu^+ \mu^-$ & 0.02   \\
$H \to c\overline{c}$ & 2.9   & &  \\
 \hline
  \end{tabular}
  \caption{Summary of Higgs decay modes (BFs and resolutions) for the 125 GeV mass SM Higgs boson. For the good mass resolution channels involving a Z-boson, the resolution is only good for leptonic decays of the Z.}
  \label{tab:HiggsDecays}
\end{table}

Table~\ref{tab:higgsStatus} shows the status of the main Higgs production and decay modes, in terms of the significance of the measured signal. 
It shows that all of the main production and decay modes have been established, although many of these were only observed in the last year.

\begin{table}[tbhp]
   \centering
  \begin{tabular}{|l|c|c|c|c|c|c|c|c|}
  \hline
 & $\gamma \gamma$ & $ZZ^\ast$ & $WW^\ast$ & $b\overline{b}$ & $c\overline{c}$ & $\tau^+\tau^-$ & $\mu^+ \mu^-$ & \bf{Combined} \\
  \hline
ggF & Obs. & Obs. & Obs. & - & - & UL & UL & Obs. \\
VBF & UL & UL & UL & UL & - & Evid. & UL & Obs. \\ 
VH & UL & UL  & UL & Obs. & UL & - & - & Obs. \\
ttH & Evid. & UL & Evid. & UL & - & Evid. & - & Obs. \\
\hline
\bf{Combined} & Obs. & Obs. & Obs. & Obs. & UL & Obs. & UL & - \\ 
\hline
 \hline
  \end{tabular}
  \caption{Status of the measured significance for the main Higgs production and decay modes. Here Obs./Evid. means the significance is at the level of an observation/evidence, UL stands for 'Upper Limit' and '-' implies this mode has not been studied. }
  \label{tab:higgsStatus}
\end{table}

\begin{figure}[htb]
   \centering
   \includegraphics[width=75mm]{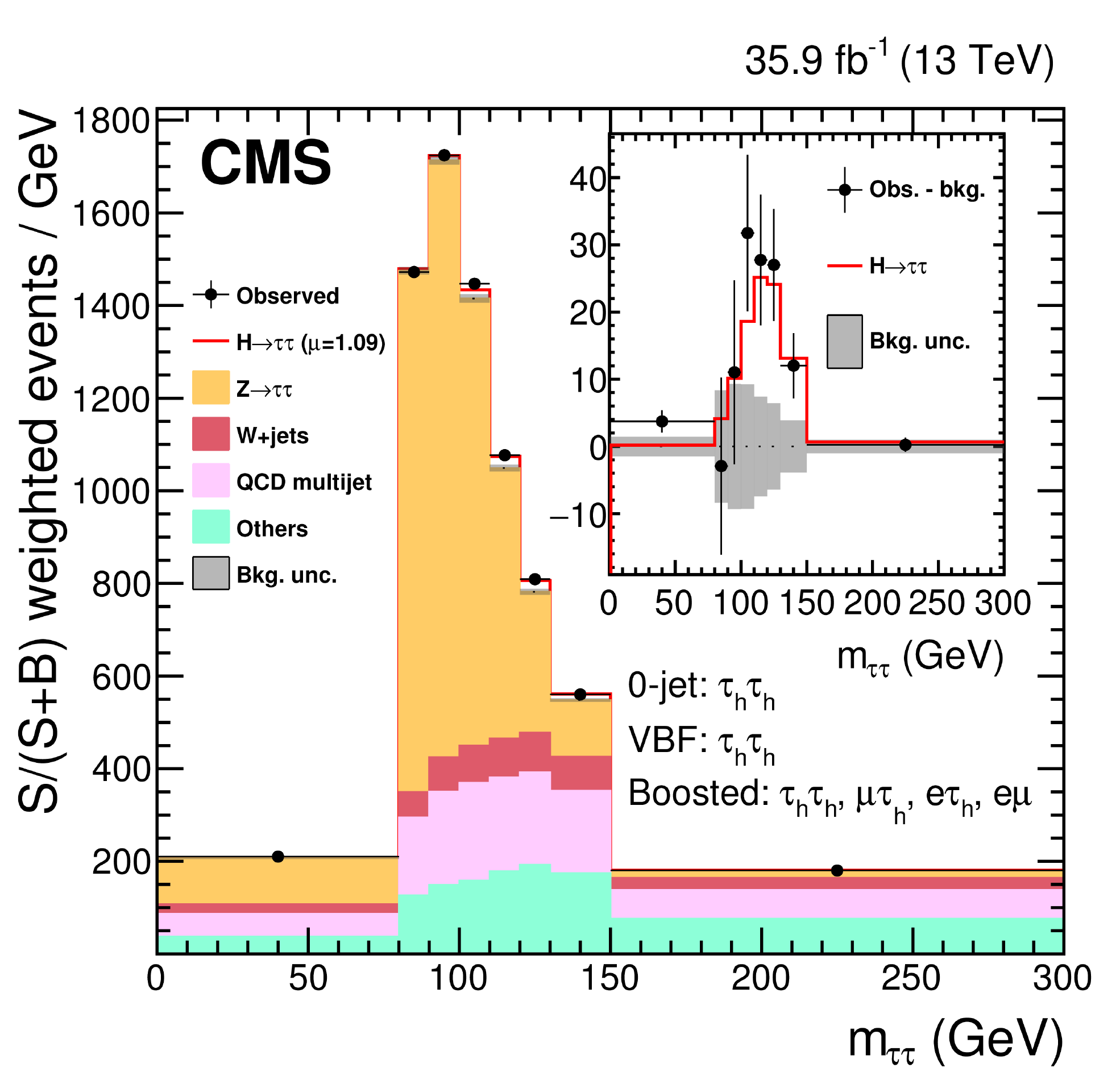}
   \includegraphics[width=75mm]{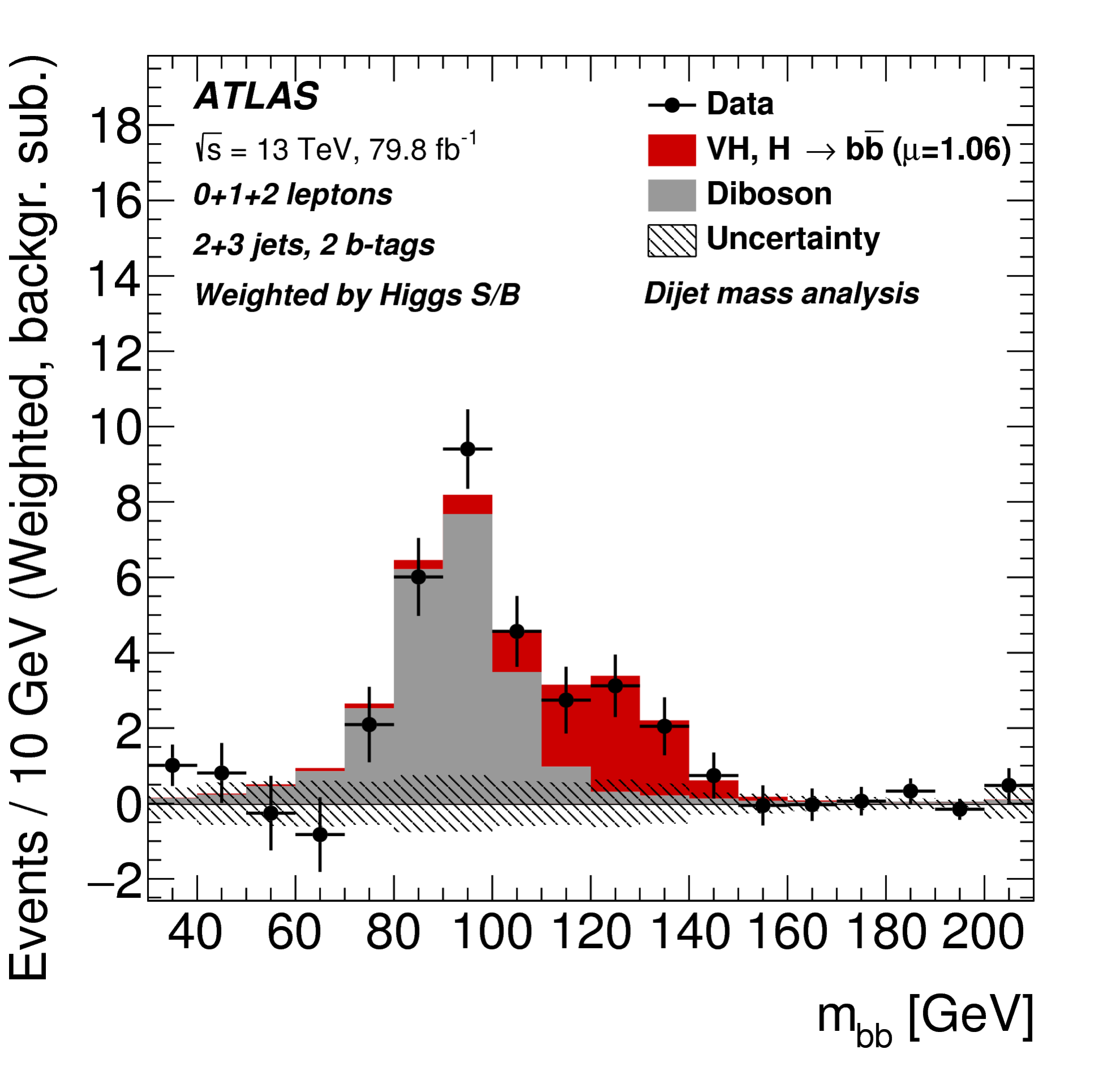}
   \caption{(left) The di-tau mass distribution from the CMS $H \to \tau^+ \tau^-$ analysis; (right) The di-$b$-jet mass distribution from the ATLAS $H \to b \overline{b}$ analysis.}
   \label{fig:higgsTauTau-bb}
\end{figure}

The Higgs coupling to fermions was established with the observation of the $H \to \tau^+ \tau^-$ decay. 
The analysis selects events with two $\tau$s (either can decay hadronically or leptonically), and uses selections targeting either VBF Higgs 
production or high-\pT\ ggF Higgs production to reduce the backgrounds. The main background is $Z \to \tau^+ \tau^-$ which has the same final state, 
with $\approx$1000$\times$ higher cross-section and with a similar di-$\tau$ mass (the mass resolution is not sufficient to be able to resolve the two 
processes). Figure~\ref{fig:higgsTauTau-bb} shows the di-$\tau$ mass distribution from the Run-2 CMS analysis~\cite{cms-tautau} where a tiny signal can be seen on top of the 
large $Z \to \tau \tau$ background. The analysis measured the $H \to \tau \tau$ rate to be compatible with the SM expectation with a precision of 
$\approx$30\%, corresponding to a 5.9$\sigma$ observation of the process. Searches for $H \to \mu^+ \mu^-$ have found no evidence of a signal (as expected in 
the SM for the current dataset), which when combined with the $H \to \tau^+ \tau^-$ result, demonstrates the Higgs couplings do not obey lepton flavour conservation.

The Higgs coupling to quarks was established with the observation of $H \to b \overline{b}$. Although this has the largest Higgs decay BF, it is experimentally 
challenging due to the large background from QCD $b \overline{b}$ production, and the poor di-$b$-jet mass resolution. In order to reduce the 
background and to trigger on the events, the analysis targets VH production where V is a $Z$ or $W$ boson decaying leptonically, so the final 
state can have 0-leptons (but large missing transverse momentum (MET) from the $Z \to \nu \overline{\nu}$ decay), 1-lepton, or 2-leptons, and 
selects two $b$-jets with a mass close to the Higgs 
mass. As seen in Figure~\ref{fig:higgsTauTau-bb}, which shows the ATLAS analysis~\cite{atlas-hbb}, the $VZ$, $Z\to b \overline{b}$ decay acts as an important validation of the analysis. 
This has the same final state, with a similar rate. The figure shows that $VZ$ is observed with the expected rate (grey), and the $H \to b \overline{b}$ signal 
can be seen as a high mass shoulder on the $Z$ peak. The analysis finds the expected SM rate with a precision of $\approx$30\%. 

The Higgs is too light to decay to top quarks, so the top-Higgs (tH) coupling can only be directly probed through ttH production. In the SM ggF 
production is dominated by a top-quark in the ggF loop, and so the tH coupling can also be extracted indirectly from ggF production rates. The 
direct and indirect measurement of the coupling then allow to constrain possible new-particles that could enter the ggF loop. Within the 
current precision the direct and indirect measurements of the coupling are compatible.

With the increased luminosity at the HL-LHC the Higgs physics goals are:
\begin{itemize}
\item Improve the precision on the Higgs couplings to the few-\% level (where they can be sensitive to BSM effects);
\item Establish the coupling to 2nd generation fermions through the $H \to \mu^+ \mu^-$ and $H \to c \overline{c}$ decays;
\item Improve the constraints on forbidden Higgs decays such as $H \to $ invisible and lepton-flavour violating Higgs decays;
\item Make more precise differential measurements of Higgs production, in more extreme regions of phase-space) which can be sensitive to New Physics;
\item Observe the very rare di-Higgs production process.
\end{itemize}
Studying di-Higgs production is needed to understand the Higgs self-coupling, and to probe the Higgs potential term of the SM Lagrangian. However, 
it is doubtful that this will be possible at the HL-LHC. Current projections~\cite{hh-hl-lhc-cms}~\cite{hh-hl-lhc-atlas}  suggest that evidence for 
di-Higgs production can be achieved by combining the ATLAS and CMS HL-LHC results.

\subsection{Searches for Physics beyond the Standard Model}
One of the primary goals of the LHC is to search for the direct production of Beyond the Standard Model (BSM) physics. ATLAS and CMS have carried out a huge number of searches, 
but to date no significant excess of events over the SM expectation has been observed. A few example searches are discussed below.

\begin{figure}[htb]
   \centering
   \includegraphics[width=75mm]{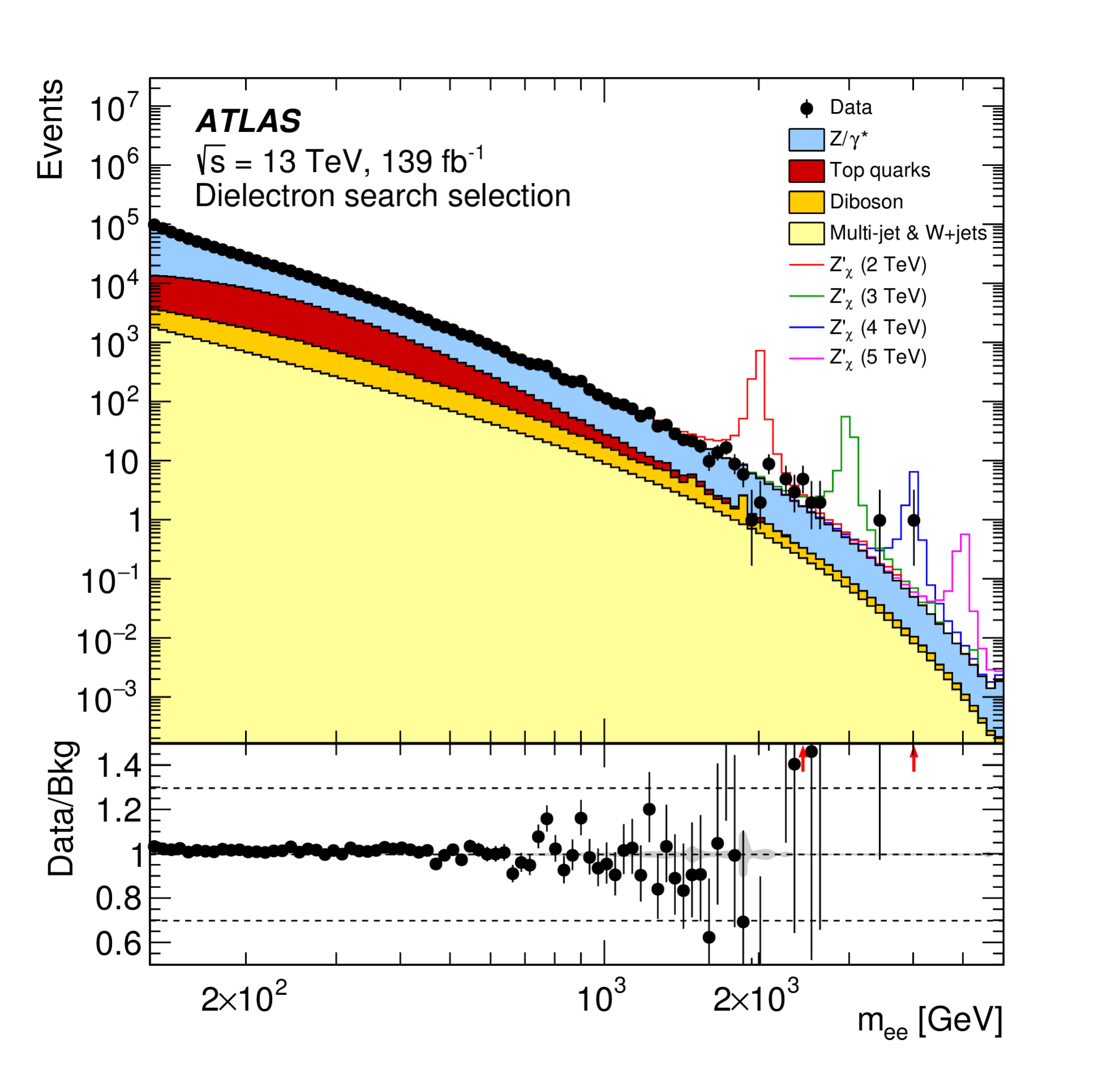}
   \includegraphics[width=75mm]{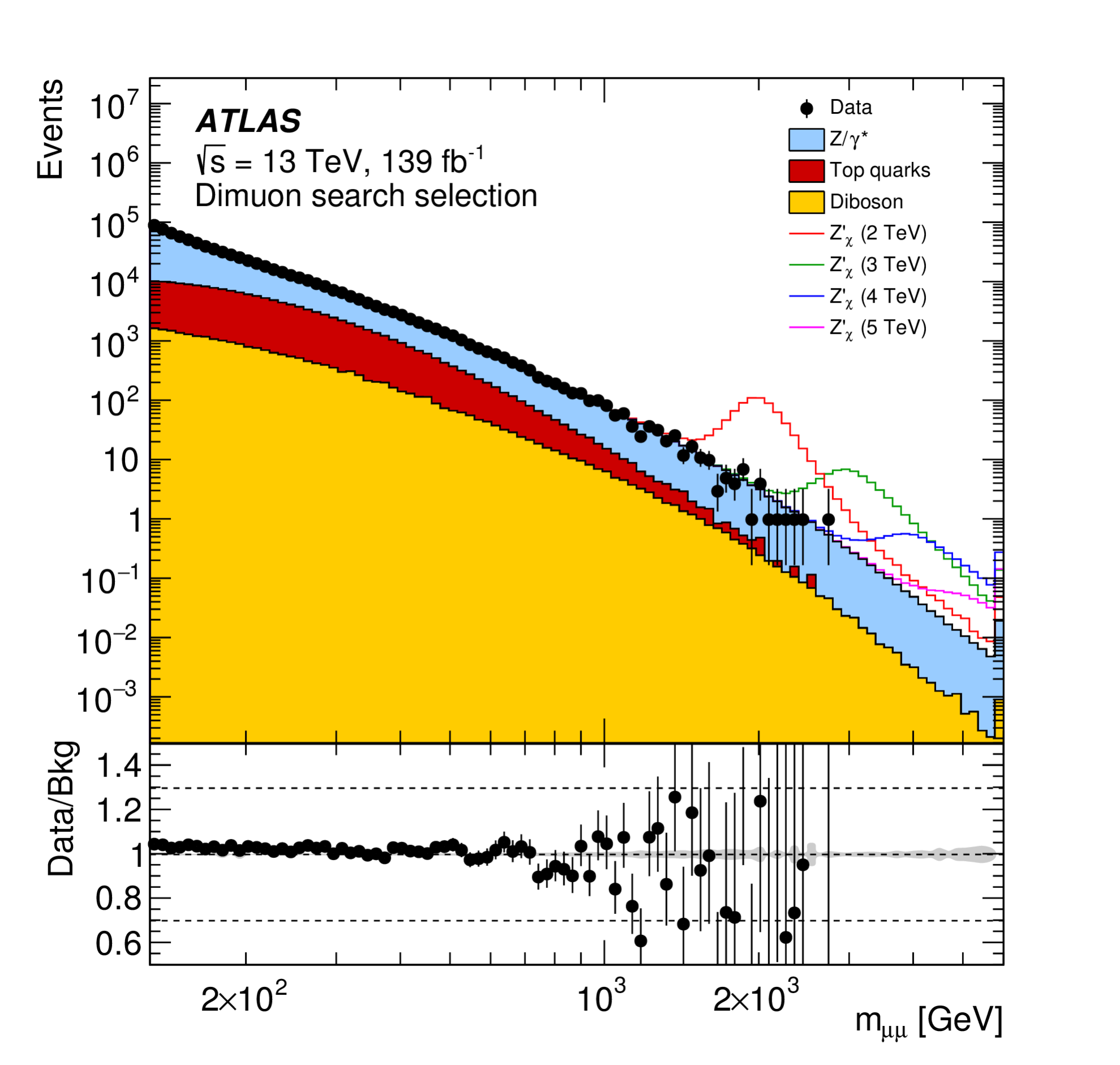}
   \caption{The di-electron and di-muon mass distributions from the ATLAS $Z'$ search.}
   \label{fig:Zprime}
\end{figure}

A search for a new Gauge boson ($Z'$) that is similar to the SM $Z$ boson but with much higher mass, looks for an excess of events in the di-lepton mass spectra 
at high mass. Figure~\ref{fig:Zprime} shows the di-electron and di-muon mass distributions from the ATLAS search~\cite{atlas-zprime}. No significant deviation from the expected 
background (dominated by SM Drell-Yan production) is observed. Examples signals are shown in the figures, which show that the mass resolution is significantly 
better at high mass for electrons than for muons, as the energy resolution improves for calorimeters, but deteriorates for tracking detectors, at higher energy. 
The main experimental challenge for this search is to have good efficiency and resolution for very high transverse momentum leptons (up to 2 TeV).

At the other end of the spectrum is a search for Higgsino production where very low-momentum leptons are expected. The ATLAS search~\cite{atlas-higgsino} uses leptons with \pT\ 
down to 3 GeV (muons) and 4.5 GeV (electrons) in order to improve the sensitivity, and allows to exclude Higgsinos with masses up to 150 GeV for certain mass splittings.

\begin{figure}[htb]
   \centering
   \includegraphics[width=55mm, height=66mm]{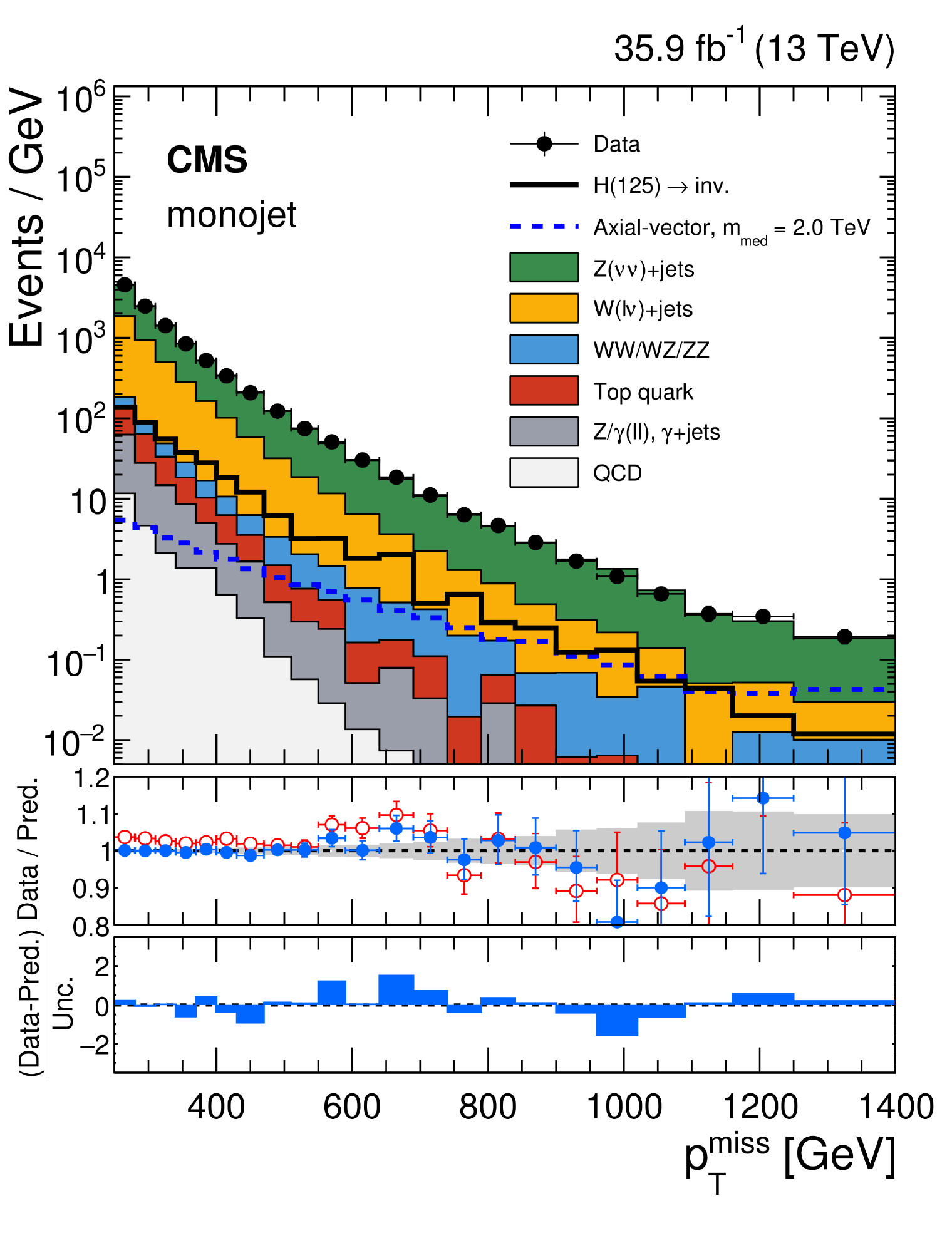}
   \includegraphics[width=50mm, height=66mm]{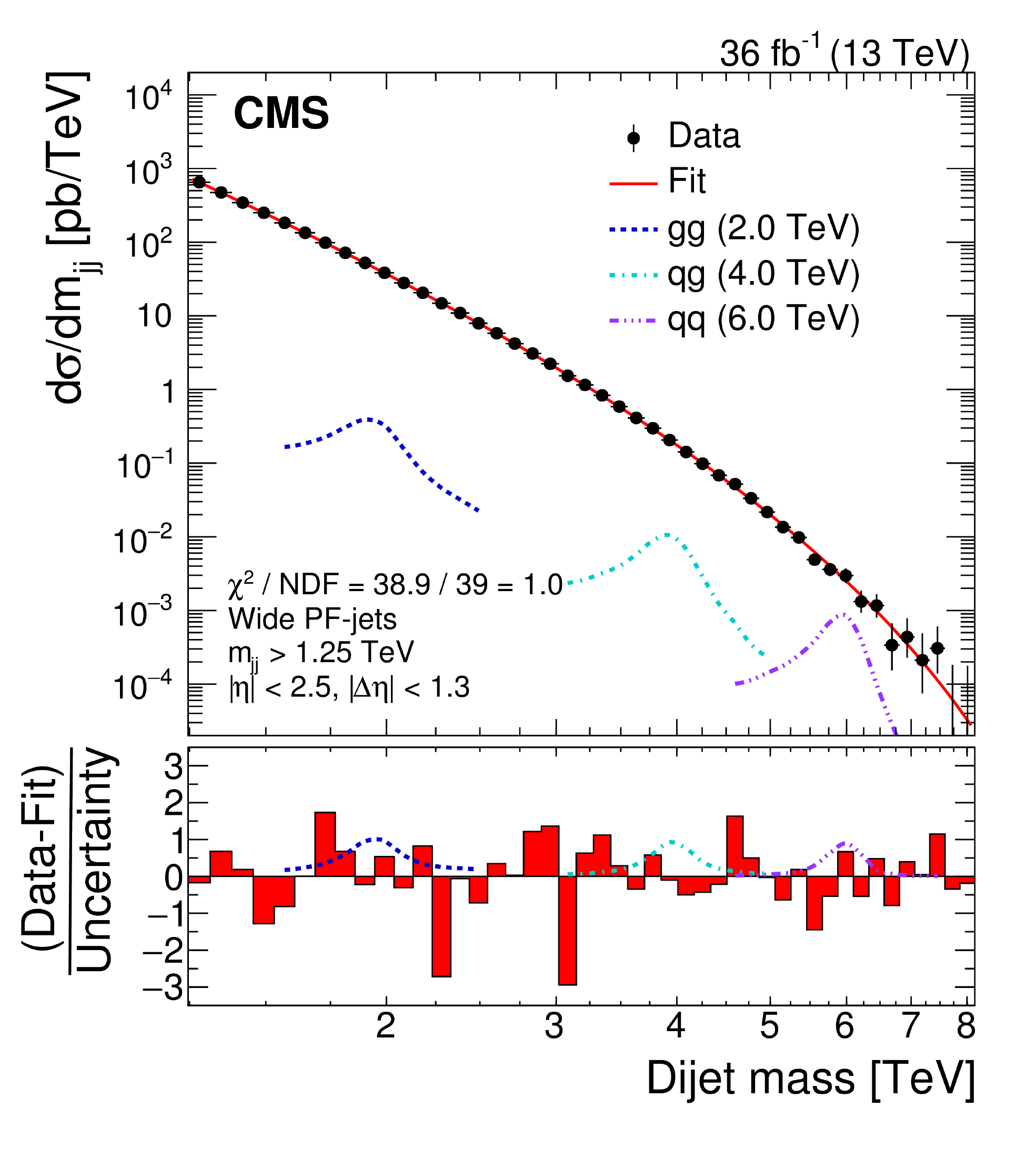}
   \includegraphics[width=50mm,height=66mm]{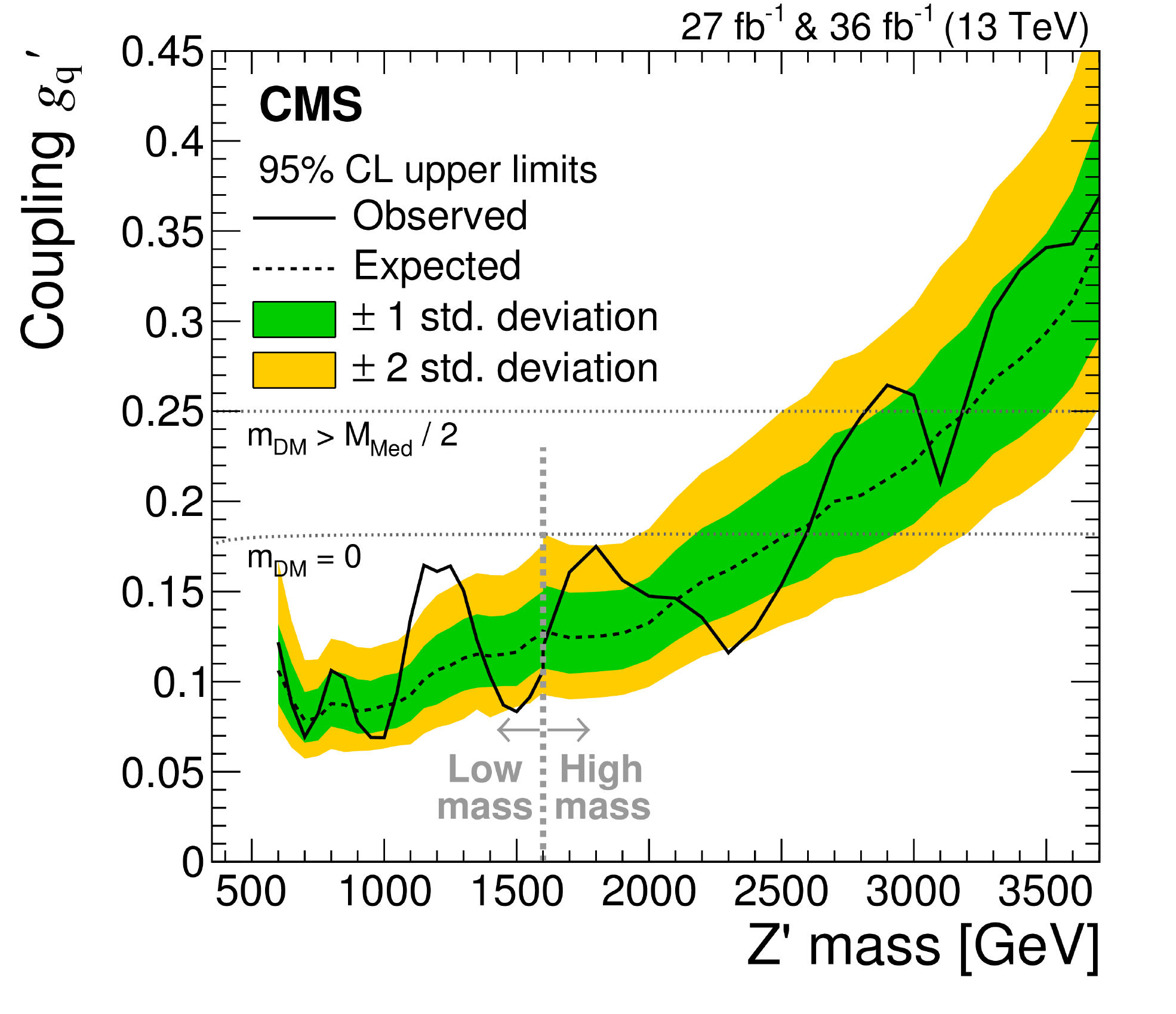}
   \caption{(left) The MET spectrum in the DM search; (middle) The di-jet mass spectra in the mediator search; (right) The exclusion limit in the search for the mediator showing 
results for both the high mass search, and the low mass search which uses the {\it Data Scouting} technique.}
   \label{fig:monojet}
\end{figure}

Searching for Dark Matter (DM) production in LHC collisions can be done by taking advantage of initial-state-radiation which can be used to tag events where DM particles are pair 
produced through an $s$-channel mediator particle but escape the detector without interacting with it. This can lead to a detector signature of a high-\pT\ jet + MET. 
Figure~\ref{fig:monojet} shows the MET spectrum for such events from the CMS search~\cite{cms-dm}, also showing the expected background, dominated by 
$Z \to \nu \overline{\nu}$ + jets ($\approx$60\%) and $W \to \ell \nu$ + jets (where the lepton is not reconstructed)  ($\approx$30\%). The signal has a slightly harder 
MET-spectra than the background, but is much smaller than the background, meaning the background needs to be controlled at the few-\% level to allow to have sensitivity. 
The background is estimated from data control regions with $Z \to \ell^+ \ell^- $ + jets, $W \to \ell \nu$ + jets and $\gamma$ + jets but accurate theoretical predictions 
are needed on the ratio of $Z$ + jets/$\gamma$ + jets and $Z$ + jets/$W$ + jets; in order to achieve the needed precision nNLO electroweak corrections need to be taken into 
account. 

As well as searching for the DM particle, we can also search for the mediator which can be produced in the LHC collisions but decay back to the SM (for example to two quarks) 
this could show up as a resonance in the di-jet mass spectra. Figure~\ref{fig:monojet} shows the di-jet mass distribution for such a CMS search~\cite{cms-datascouting}, showing a smoothly falling distribution 
with no sign of a resonance in the range 1 TeV $\to$ 8 TeV in di-jet mass. The 1 TeV lower limit in the probed mass range, comes from the trigger thresholds applied to 
the jets used in the search. Going to lower \pT -jets would increase the trigger rate leading to a too high bandwidth when reading out the detector. In order to search 
for possible resonances at lower mass a new technique called {\it Data Scouting} or {\it Trigger Level Analysis} was developed, in which just the trigger 
level jets are written out for certain triggers. These trigger level jets are much smaller than the full event data (less than 5\% of the size), and can therefore be read out at a 
much higher rate without hitting bandwidth limitations. Thus lower thresholds can be applied. This technique allows to set limits on di-jet mass resonances 
down to lower masses as can be seen in Figure~\ref{fig:monojet} (CMS analysis~\cite{cms-datascouting}).

\subsection{Precise Standard Model Measurements}
The LHC experiments carry out a large number of precise measurements of SM processes, measuring cross-sections, masses and other SM parameters. Cross-section measurements 
are normalized by the luminosity, which is measured by dedicated luminosity detectors in the experiments that are calibrated by dedicated van-der-Meer scans which 
are typically carried out each year. The precision of the luminosity measurements in ATLAS/CMS for the Run-2 dataset is an impressive $\approx$2.5\%, which is far better 
than had thought to be possible before LHC running.

An example of a very precise cross-section measurement is the $W$ and $Z$ inclusive production cross-section measurement from ATLAS~\cite{atlas-wz} with the 2011 7 TeV dataset. 
The precision is limited by systematic uncertainties, and the total experimental uncertainty is $\approx$0.5\% dominated by uncertainties related to the lepton reconstruction, 
the background (for the $W$) and theoretical modeling uncertainties (for the $Z$). The luminosity uncertainty is 1.8\%, but this cancels in ratios such as 
$\sigma(W \to e \nu) / \sigma(W \to \mu \nu)$ or $\sigma(W)/\sigma(Z)$  allowing very  precise tests of lepton flavour conservation, and parton distribution functions (PDFs).

The measurement of the $W$-boson mass by ATLAS~\cite{atlas-wamass} with a precision of 19~MeV represents one of the most precise measurements at the LHC, and has a precision equal 
to the best single-experiment measurement. The $W$ mass is a fundamental parameter of the SM, and has important sensitivity in the electroweak fit. The ATLAS analysis measures the 
mass using a template fits to the transverse mass (formed from the lepton and the reconstructed hadronic recoil), and to the lepton transverse momentum. A very precise knowledge 
of experimental effects related to lepton reconstruction and the hadronic recoil reconstruction is needed, where the later deteriorates significantly with pileup. The current 
measurement utilizes the 2011 7~TeV data set which has an average pileup of around 9. Theoretical uncertainties also play an important role, in particular related to the 
modelling of the $W$-boson \pT\ which is derived from the measured $Z$-boson \pT\ spectra, as well as from PDF uncertainties. Utilizing low-pileup data taken in 2017 and 2018 at 
13~TeV there is the prospect of improving the precision of the measurement to the 10-15~MeV level.

Measurements of the top-quark mass are carried out in a number of different channels. A recent example from CMS~\cite{cms-topmass} utilizes the lepton+jets final state to measure 
the mass using a kinematic fit (including the $W$-mass constraint on the hadronic $W$ decay) to improve the resolution and to reduce the fraction of incorrect assignments of 
jets to the two top-quarks. The dominant systematic uncertainty is related to the jet energy scale which is constrained in the fit. The final result of 
$172.25 \pm 0.08$(stat.)$ \pm 0.62$(syst.)~GeV is the most precise single measurement to date.

\subsection{Flavour Physics}
The $B_S \to \mu^+ \mu^-$ rare decay is theoretically clean, and has a large sensitivity to many new physics models (for example MSSM scenarios with large Tan $\beta$). Because of 
this there is a long history of searches for this decay starting over 30 years ago. Sensitivity to the SM branching ratio of $(3.3 \pm 0.3) \times 10^{-9}$ was reached with a 
combination between LHCb and CMS~\cite{lhcb-cms-bmm}. Despite a much smaller dataset, LHCb has the best sensitivity due to the excellent track resolution, as well as an optimized 
trigger for low \pT\ physics; CMS has better sensitivity than ATLAS due to the higher magnetic field in the inner tracker which gives a better mass resolution. Current 
measurements from all three experiments are consistent with the SM estimate with an uncertainty of between 20 - 30 \%.

\begin{figure}[htb]
   \centering
   \includegraphics[width=52mm]{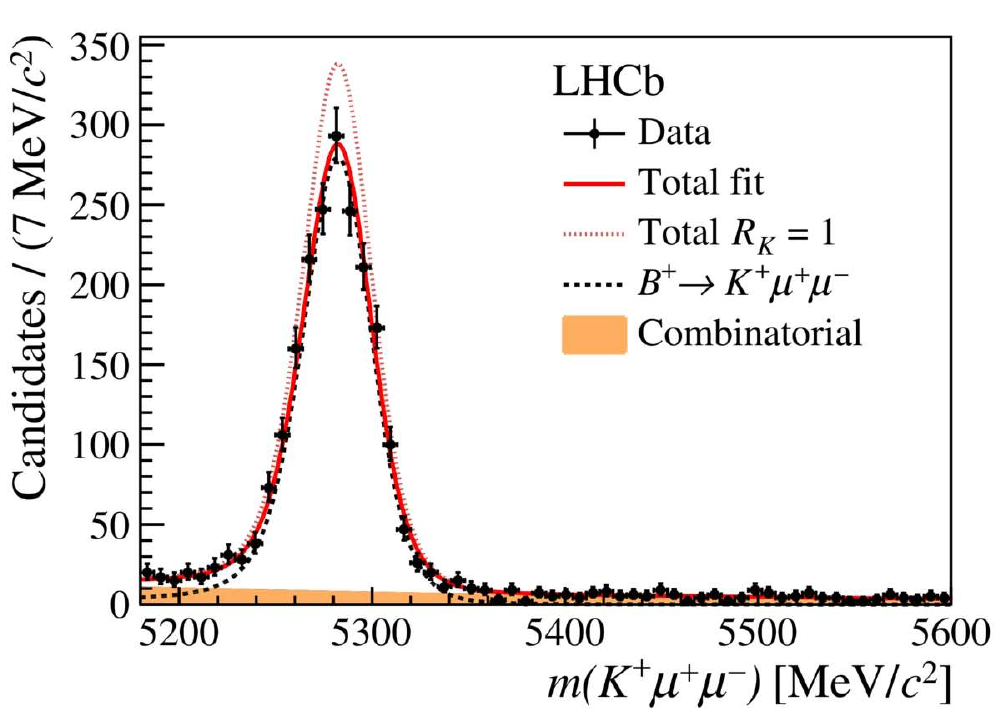} 
   \includegraphics[width=52mm]{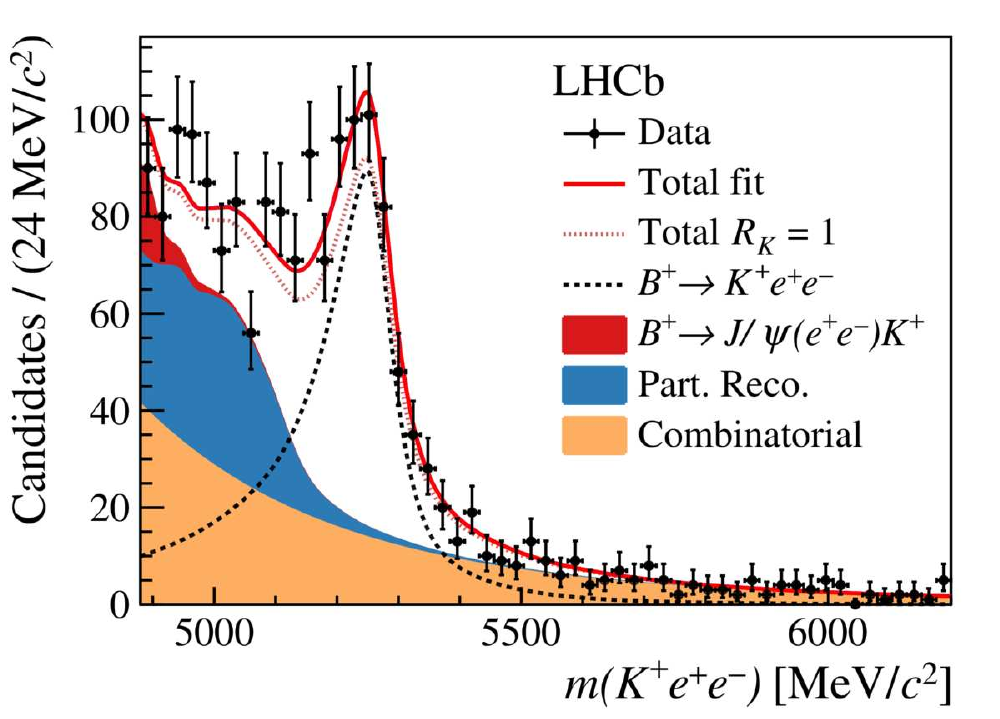}
   \includegraphics[width=52mm]{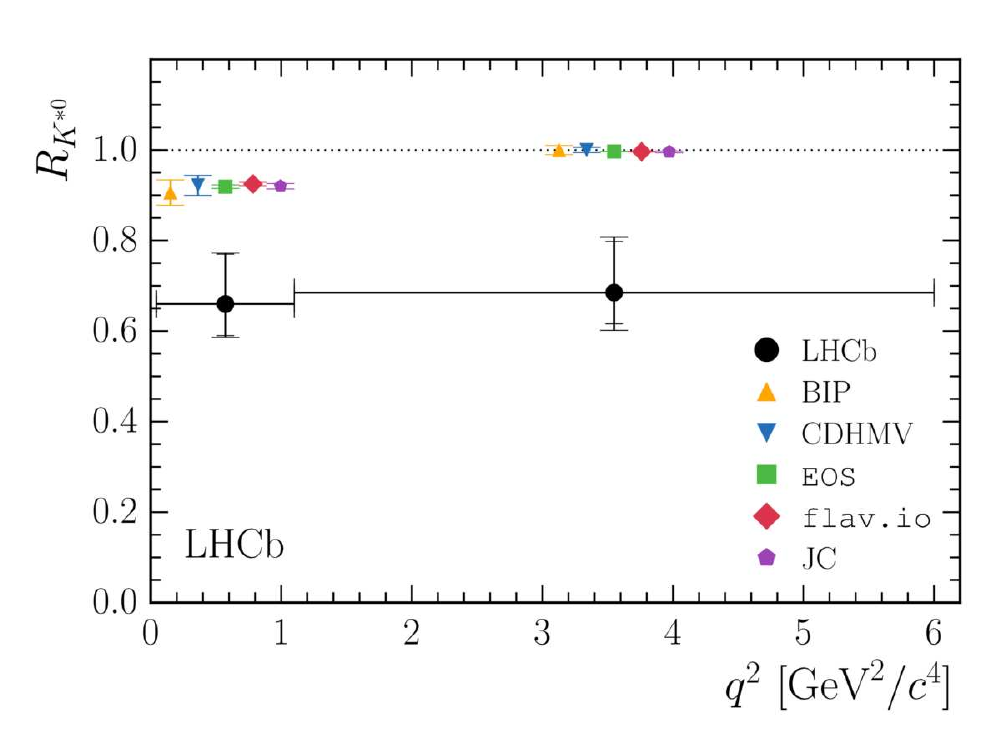}
   \caption{(left/middle) The reconstructed $B$ meson mass in the $R_K$ analysis for the  $ \mu^+ \mu^-$ / $e^+ e^-$ channels; (right) The measured $R_{K^*}$ values in two bins of 
$q^2$ (the di-lepton mass) compared with various theoretical predictions.}
   \label{fig:LFV}
\end{figure}

LHCb searches for lepton flavour violation in $B$ meson decays by measuring the ratio $R_{K^{(*)}} \equiv {\rm BF}(B \to K^{(*)} \mu^+ \mu^-) / {\rm BF}(B \to K^{(*)} e^+ e^-) $. 
In the SM this is precisely predicted and is close to unity, modulo phase-space effects. An experimental complication is that due to bremsstrahlung the mass resolution is much 
worse in the di-electron channel than the di-muon channel (as can be seen in Figure~\ref{fig:LFV}), but this is corrected for by normalizing by the measured ratio BF$(B \to K^{(*)} J/\psi(\mu^+ \mu^-))$
/ BF$(B \to K^{(*)} J/\psi(e^+ e^-))$. For $R_K$, the reconstructed $B$ meson mass distribution is shown in Figure~\ref{fig:LFV} in both the $ \mu^+ \mu^-$ and $e^+ e^-$ channels.
The measured values from the LHCb measurements~\cite{rkstar-lhcb}~\cite{rk-lhcb} are shown in Table~\ref{tab:LFV}, and for $R_{K^*}$ in Figure~\ref{fig:LFV}, and show that the 
three measurements are between 2 and 2.5 $\sigma$ lower than the SM prediction. This is currently one of the most intriguing anomalies observed at the LHC, with many theoretical 
models proposed to explain the results. More data and measurements from Belle-2 should shed further light onto the situation.

\begin{table}[tbhp]
   \centering
  \begin{tabular}{|l|c|c|c|}
  \hline
Measurement & Dataset & Measured Value & Compatibility with SM \\
  \hline
$R_K$ & Run 1 + Run 2 & $0.85^{+0.06}_{-0.05} \pm 0.015$ & 2.5$\sigma$ \\
$R_{K^*}$ low-q$^2$ &Run 1 & $0.66^{+0.11}_{-0.07} \pm 0.03$ & 2.2$\sigma$ \\
$R_{K^*}$ high-q$^2$ &Run 1 & $0.69^{+0.11}_{-0.07} \pm 0.05$ & 2.4$\sigma$ \\
 \hline
  \end{tabular}
  \caption{LHCb results on lepton flavour violation measurements $R_K$ and $R_{K^*}$, where the latter is measured in two regions of $q^2$ (the di-lepton mass).}
  \label{tab:LFV}
\end{table}

\section{Summary}
The LHC machine and the experiments performed extremely well in Run-2. A large and high-quality dataset was produced by the experiments leading to a huge number of 
physics results. A leading challenge for the experiments was the high pileup in the data, but they have coped very well with this situation.
 
The large dataset has allowed a more and more precise probing of the Higgs boson, where all major production modes, and decay channels accessible at the LHC have been established. 
A huge number of direct searches for BSM physics have been carried out, with no significant excess of events over the SM prediction observed, such that increasingly stringent 
exclusion limits have been set on BSM model parameters. 
In addition, the experiments have been able to make very precise measurements of cross-sections and SM parameters, as well as measuring extremely rare processes, but again no discrepancy 
with the SM expectations have been observed. 
An intriguing set of results from lepton-flavour violation measurements by LHCb show a 2 - 2.5 standard deviation discrepancy with the SM in a few channels and $q^2$-bins. 

The increased dataset that will be produced with Run-3, and then with the HL-LHC, along with the upgraded detector functionality, and innovations in triggering, reconstruction and 
physics analysis will allow to probe further the SM in the coming years.

\end{document}